\documentclass[10pt,a4paper]{article} 
\usepackage{lipsum}
\usepackage{url}
\usepackage[nochapters]{classicthesis} 
\usepackage{graphicx}
\usepackage{hyperref}
\usepackage{color, soul}
\usepackage[labelfont=bf]{caption}
\usepackage{placeins}
\usepackage[margin=1.25in]{geometry}
\usepackage{amsmath}
\usepackage{esvect}

\begin{document}

    \pagestyle{plain}
    \font\myfont=cmr8 at 8pt
    \title{\normalfont{The DynaSig-ML Python package: automated learning of biomolecular dynamics-function relationships}}
    \author{Olivier Mailhot$^{1-4}$ \and François Major$^{2,3}$ \and Rafael Najmanovich$^{4,*}$}
    \date{%
    \myfont$^1$Department of Biochemistry and Molecular Medicine, Université de Montréal, Montreal, Canada\\%
    $^2$Department of Computer Science and Operations Research, Université de Montréal, Montreal, Canada\\%
    $^3$Institute for Research in Immunology and Cancer, Université de Montréal, Montreal, Canada\\%
    $^4$Department of Pharmacology and Physiology, Université de Montréal, Montreal, Canada\\%
    $^*$To whom correspondence should be addressed.\\%
    \today
}

    \maketitle

    \begin{abstract}
        \noindent \textbf{Summary:} The DynaSig-ML ("Dynamical Signatures - Machine Learning")
    Python package allows the efficient, user-friendly exploration of 3D dynamics-function relationships
    in biomolecules, using datasets of experimental measures from large numbers of sequence variants.
    The DynaSig-ML package is built around the Elastic Network Contact Model (ENCoM), the first and only
    sequence-sensitive coarse-grained NMA model, which is used to generate the input Dynamical
    Signatures. Starting from \textit{in silico} mutated structures, the whole pipeline can be run
    with just a few lines of Python and modest computational resources. The compute-intensive steps can
    also easily be parallelized in the case of either large biomolecules or vast amounts of sequence
    variants. As an example application, we use the DynaSig-ML package to predict the evolutionary
    fitness of the bacterial enzyme VIM-2 lactamase from deep mutational scan data.\\%

        \noindent \textbf{Availability and implementation:} DynaSig-ML is open source software available
        at \url{https://github.com/gregorpatof/dynasigml_package}\\%
        
        \noindent \textbf{Contact:} \href{mailto:rafael.najmanovich@umontreal.ca}{rafael.najmanovich@umontreal.ca}

    \end{abstract}

    \newpage

    \section{Introduction}

    The Elastic Network Contact Model (ENCoM) is the only sequence-sensitive coarse-grained normal
    mode analysis model \cite{Frappier_2014}. Its sequence sensitivity enables its use to predict
    the impact of sequence variants on biomolecular function through changes in predicted stability
    \cite{Frappier_2014b} and dynamics \cite{Teruel_2021}. We recently extended ENCoM to work on RNA
    molecules and predicted microRNA maturation efficiency from a dataset of experimentally measured
    maturation efficiencies of over 26 000 sequence variants \cite{Mailhot_2022}. To do so, the ENCoM Dynamical
    Signatures, which are vectors of predicted structural fluctuations at every position in the system, were used as input
    variables in a LASSO multiple linear regression model \cite{Tibshirani_1996} to predict maturation
    efficiency. To our knowledge, this coupling of coarse-grained normal mode analysis to machine
    learning in order to predict biomolecular function is the first of its kind. Furthermore, it can
    be applied to any biomolecule for which there exist experimental data linking perturbations
    (such as mutations or ligand binding) to function. Indeed, ENCoM is currently applicable to
    proteins, nucleic acids, small molecules and their complexes \cite{Mailhot_2021}. Here we
    present the DynaSig-ML ("Dynamical Signatures - Machine Learning") Python package, which allows the
    implementation and automated replication of that novel protocol. As an example application, we
    apply DynaSig-ML to predict enzymatic efficiencies of VIM-2 lactamase sequence variants, starting from mutagenesis
    data. DynaSig-ML automatically computes the ENCoM Dynamical Signatures from a list of perturbed
    structures (mutations or ligand binding), stores them as lightweight serialized files, and can then
    be used to train simple machine learning algorithms using the Dynamical Signatures. The first algorithm is LASSO
    regression, which allows the mapping of the learned coefficients on the studied structure
    (automatically accomplished by DynaSig-ML plus two simple PyMOL \cite{Delano_2002} commands).
    As these coefficients represent the relationship between flexibility change at specific positions
    and the predicted functional property, this mapping can be used to drive new biological hypotheses.
    The second machine learning model implemented is the multilayer perceptron (MLP), a
    type of feedforward neural network \cite{Murtagh_1991}. MLPs can learn complex relationships
    between the input variables and are thus more powerful than LASSO regression, however it is not
    possible to map the learned patterns back on the structure because of the MLP's complexity and
    absence of linear independence between input variables. DynaSig-ML automatically generates graphs
    showing testing performance. Each of the
    necessary steps to apply DynaSig-ML is documented online as part of a step by step tutorial
    (\url{https://dynasigml.readthedocs.io}).

\section{Implementation}

    DynaSig-ML runs the ENCoM model within NRGTEN, another user-friendly, extensively
    documented Python package \cite{Mailhot_2021}. The machine learning models are implemented using
    the scikit-learn Python package \cite{Pedregosa_2011}. The numerical computing is accomplished
    by NumPy \cite{Oliphant_2006} and the performance graphs are generated with matplotlib
    \cite{Hunter_2007}, making these four packages the only dependencies of DynaSig-ML.

\section{VIM-2 lactamase example}

\begin{figure}[bth]
    \includegraphics[width=\linewidth]{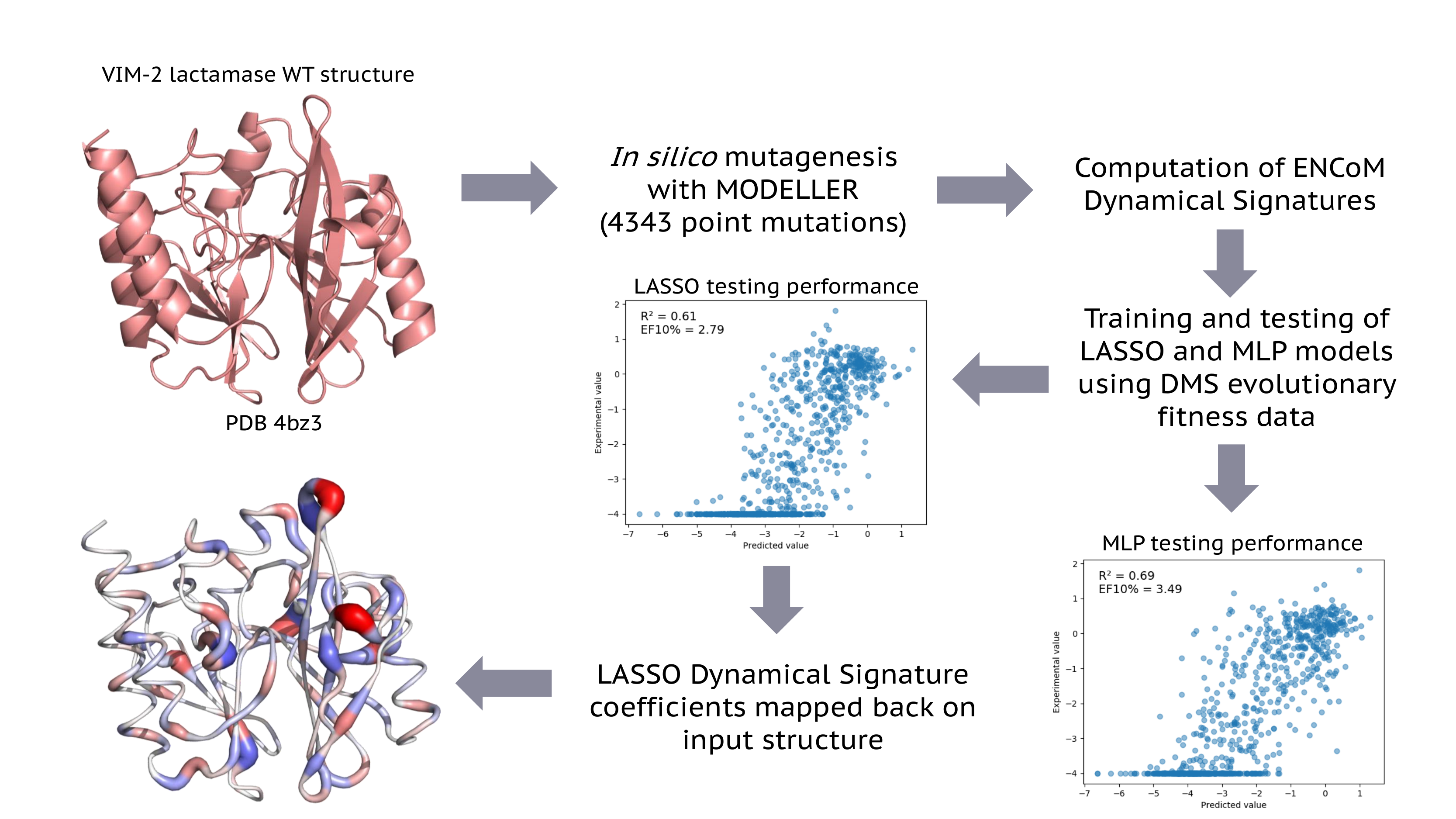} \caption{\textbf{ENCoM-DynaSig-ML pipeline
    applied to VIM-2 lactamase deep mutational scan data.} The crystal structure of VIM-2 lactamase is
    used as a template to perform the 4343 point mutations with experimental fitness data using the
    MODELLER software \cite{Webb_2014}, all subsequent steps are performed using DynaSig-ML. For
    each of the \textit{in silico} variants, a Dynamical
    Signature is computed with ENCoM. LASSO regression and multilayer perceptron models are trained
    using as input variables the Dynamical Signatures and the structural attributes used by Chen
    \textit{et al.} as part of the VIM-2 deep mutational scan study \cite{Chen_2020}. In the case of the
    LASSO regression model, the independence of the input variables allows the mapping of the learned
    coefficients back on the VIM-2 structure. The color gradient represents each coefficient, from
    blue for negative coefficients, to white for null coefficients and red for positive coefficients.
    The largest absolute value coefficient will have the brightest color. The sign of a coefficient captures
    the nature of the relationship between flexibility changes at that position and the experimental property of interest
    (in this case, evolutionary fitness). Negative coefficients mean that rigidification of the position
    leads to higher fitness, while positive coefficients mean that softening of that position leads
    to higher fitness. The thickness of the cartoon represents the absolute value of the coefficients,
    i.e. their relative importance in the model.}\label{fig:dynasigml}
\end{figure}

    In order to illustrate a typical use case of DynaSig-ML, we applied it to study
    dynamics-function relationships from deep mutational scan (DMS) data on the VIM-2 lactamase
    enzyme \cite{Chen_2020}. VIM-2 (Verona integron-encoded metallo-$\beta$-lactamase 2) is a
    bacterial enzyme capable of degrading $\beta$-lactam antibiotics, and represents a major source
    of worldwide antibiotic resistance \cite{Christopeit_2016}. The DMS dataset used measured
    bacterial fitness for every VIM-2 sequence variant under various concentrations of antibiotics
    \cite{Chen_2020}. For this application, we use the fitness under the maximal concentration of
    ampicillin (128$\mu$g/mL) at 37 degrees Celsius as the property the machine learning models
    try to predict from the Dynamical Signatures. \autoref{fig:dynasigml} illustrates the whole
    protocol used to start from the PDB \cite{Sussman_1998} structure of VIM-2 lactamase \cite{Brem_2016}, train the
    machine learning models, test their performance and map the LASSO coefficients back on the VIM-2
    structure.

\FloatBarrier

    Interestingly, the Dynamical Signatures exhibit good complementarity to the other attributes
    used by Chen \textit{et al.}, which are the mean $\Delta\Delta$G of folding calculated with
    Rosetta \cite{Rohl_2004}, the change in
    accessible surface area, and a vector of length 40 which specifies what are the starting and
    mutated amino acid identities for the point mutation. The authors report a training coefficient of
    determination of $\text{R}^2=0.55$ when fitting a linear model with these input variables to
    all the available data. They do not report testing performance, so this $0.55~\text{R}^2$ can be
    seen as an upper bound for the performance of a linear model based on these properties. Since the
    dataset contains point mutations only, there can be no sequence redundancy between the
    training and testing set. For this reason, we generated a random 80/20 train/test split for this
    application. The exact variants randomly picked for the testing set are available in the GitHub
    repository that accompanies the online DynaSig-ML tutorial
    (\url{https://github.com/gregorpatof/dynasigml_vim2_example}). When combining the Chen \textit{et al.}
    attributes and Dynamical Signatures, we obtain LASSO and MLP models reaching respective testing performances of
    $\text{R}^2=0.61$ and $\text{R}^2=0.69$. The enrichment factors at 10\%, which are values ranging from
    0 to 10 characterizing the relative proportion of the top 10\% measured values in the top 10\%
    predicted values, are 2.79 and 3.49 for the LASSO and MLP models respectively. For a more in-depth
    analysis of performance including models trained with the Dynamical Signatures alone and static
    predictors alone, see Supplementary Information.

\section{Conclusions}

    In conclusion, the DynaSig-ML Python package allows the fast and user-friendly exploration of
    dynamics-function relationships in biomolecules. It uses the ENCoM model, the first and only
    sequence-sensitive coarse-grained normal mode analysis model, to automatically compute
    Dynamical Signatures from structures in PDB format, stores them as lightweight
    serialized Python objects, and automatically trains and tests LASSO regression and MLP models
    to predict experimental measures. Moreover, it automatically generates
    performance graphs and maps the LASSO coefficients back on the input PDB structure. A detailed
    online tutorial is available to replicate the VIM-2 deep mutational scan application presented
    here (\url{https://dynasigml.readthedocs.io}).

    \section{Acknowledgements and funding}

    RJN is part of PROTEO (the Québec network for research on protein function, structure and engineering).

    This work was supported by Natural Sciences and Engineering Research Council of Canada (NSERC) Discovery program grants
    (FM and RN); Genome Canada and Genome Quebec (RN); Compute Canada (RN) and Canadian Institutes of Health Research (CIHR) (FM, grant number MOP-93679). OM is the recipient
    of a Fonds de Recherche du Québec—Nature et Technologies (FRQ-NT) Doctorate's scholarship; and a Faculté des
    Études Supérieures et Postdoctorales de l'Université de Montréal scholarship for direct passage to the PhD.
    
    Conflict of Interest: none declared.

    \addtocontents{toc}{\protect\vspace{\beforebibskip}}
    \addcontentsline{toc}{section}{\refname}
    \bibliographystyle{unsrt}
    \bibliography{used_refs}

\end{document}


\pagestyle{plain}
    \font\myfont=cmr8 at 8pt
    \title{\normalfont{Supplementary information for: The DynaSig-ML Python package:
                       automated learning of biomolecular dynamics-function relationships}}
    \author{Olivier Mailhot$^{1-4}$ \and François Major$^{2,3}$ \and Rafael Najmanovich$^{4,*}$}
    \date{%
    \myfont$^1$Department of Biochemistry and Molecular Medicine, Université de Montréal, Montreal, Canada\\%
    $^2$Department of Computer Science and Operations Research, Université de Montréal, Montreal, Canada\\%
    $^3$Institute for Research in Immunology and Cancer, Université de Montréal, Montreal, Canada\\%
    $^4$Department of Pharmacology and Physiology, Université de Montréal, Montreal, Canada\\%
    $^*$To whom correspondence should be addressed.\\%
    \today
}

    \maketitle

        \begin{center} \noindent Contact: \href{mailto:rafael.najmanovich@umontreal.ca}{rafael.najmanovich@umontreal.ca}
        \end{center}

    \section{Supplementary Information}

    \suppAll~ reports the testing performance of LASSO and MLP models when trained using the ENCoM Dynamical
    Signatures combined to the static predictors from
    Chen \textit{et al.} \cite{Chen_2020}, which are the Rosetta $\Delta\Delta$G of folding, the change in solvent
    accessible surface area and
    40 variables describing the starting and ending amino acid for the mutation (20 possibilities for each). In
    order to investigate the contributions of Dynamical Signatures and static descriptors alone, we also trained
    LASSO and MLP models using either one of these sets of variables alone. The testing performances are shown in
    \suppDynaSig~ and \suppStatic. Interestingly, while the testing R$^2$ is lower for the DynaSig
    models (0.24 vs 0.52 for LASSO, 0.46 vs 0.58 for MLP), both EF10\% are identical (2.79 for LASSO and 3.26 for MLP). This led us to hypothesize that
    the models might be enriching the same variants as their top predictions, however it is not the case.
    \suppTabTopTen~ lists all variants that in the top 10\% for any of the tested models or for the experimentally
    measured fitness. Surprisingly, despite identical EF10\% performances, the DynaSig and static models have few top
    predictions in common (14 out of 87 for the LASSO models, 29 out of 87 for the MLP models). This discrepancy in
    ranking the variants might explain the good complementarity of the two sets of features, especially when it comes
    to predictive R$^2$. Another striking finding is that the DynaSig models seem to benefit more from the power of the
    MLP, as R$^2$ almost doubles going from LASSO to MLP. This big gain in performance might be explained by the
    intrisincally nonlinear relationships between flexibility changes at different positions (the whole protein is a
    coupled system), which cannot be captured by linear regression.

    When it comes to predicting experimental fitness for variants containing more than one mutation, the starting and
    ending amino acid vector used by Chen \textit{et al.} cannot be used. However, from a biomolecular engineering
    point of view, variants containing multiple mutations are where computational methods really shine as the number
    of possible variants grows exponentially with the number of mutations. The Dynamical Signatures however can be computed
    for any sequence variant as long as the assumption that the equilibrium structure does not change considerably holds.
    We have recently predicted maturation efficiency for miR-125a sequence variants containing up to 6 mutations
    \cite{Mailhot_2022}. In order to investigate what performance one could expect using only the static properties
    generalizable to variants with multiple mutations, we trained LASSO and MLP models using only the accessible solvent
    area change and predicted $\Delta\Delta$G of folding. The testing performances are reported in \suppASAddG~
    and while the R$^2$ coefficients are fair (0.42 and 0.45 for LASSO and MLP), the EF10\% values are lower than with
    all other predictors used at 1.86 for LASSO and 2.67 for MLP. This illustrates the advantage of the Dynamical
    Signatures, as they are generalizable to multiple mutations and perform on par with the full static descriptors
    when it comes to EF10\%.

    Both training and testing R$^2$  and EF10\% are provided in \suppTabSummary~ as a means of quick comparison between the
    different models. We obtain similar training R$^2$ as Chen \textit{et al.} using their static descriptors
    (0.54 for our implementation, 0.55 is what was reported), and the combination of Dynamical Signatures and static descriptors reaches training R$^2$ of 0.64 for LASSO and 0.96 for MLP.

\begin{figure}[bth]
    \includegraphics[width=\linewidth]{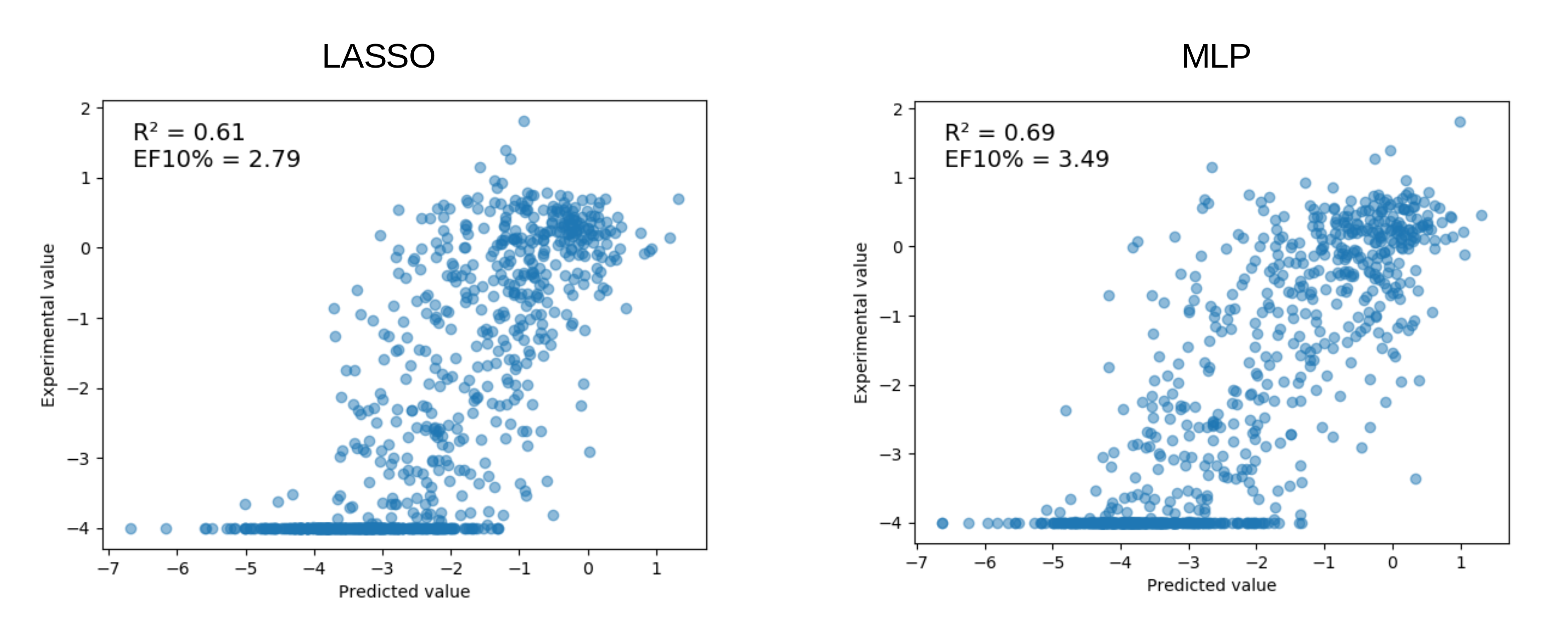}
    \caption{\textbf{Testing performance using all static descriptors and Dynamical Signatures.}}\label{fig:all} 
\end{figure}

\begin{figure}[bth]
    \includegraphics[width=\linewidth]{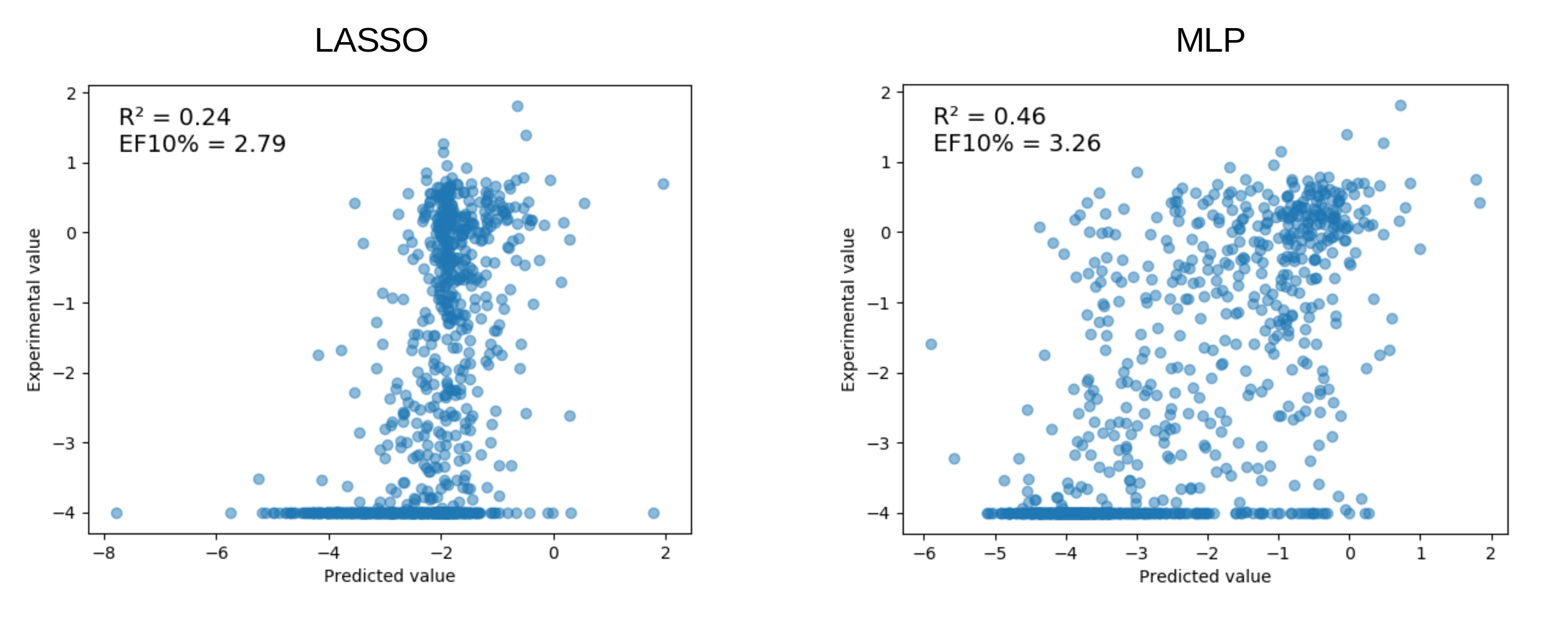}
    \caption{\textbf{Testing performance using only Dynamical Signatures.}}\label{fig:dynasig} 
\end{figure}

\begin{figure}[bth]
    \includegraphics[width=\linewidth]{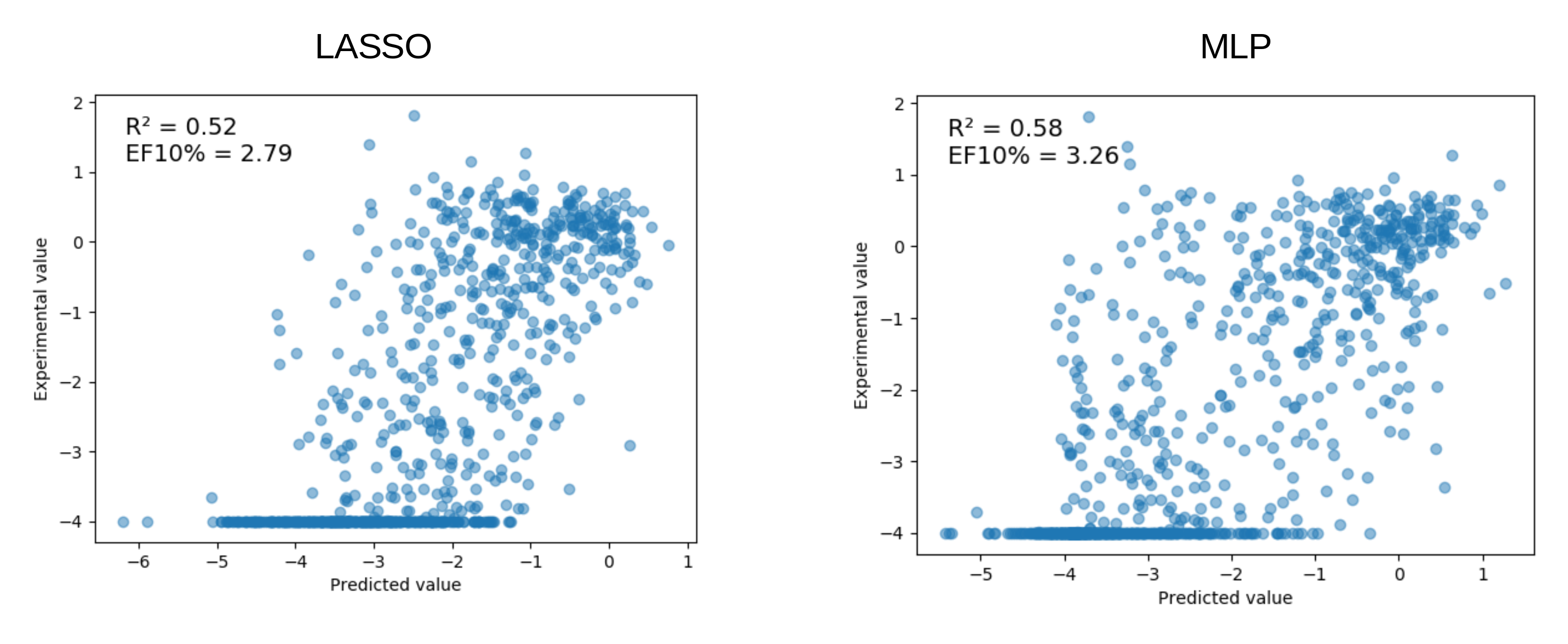}
    \caption{\textbf{Testing performance using only static descriptors.}}\label{fig:static} 
\end{figure}  

\begin{figure}[bth]
    \includegraphics[width=\linewidth]{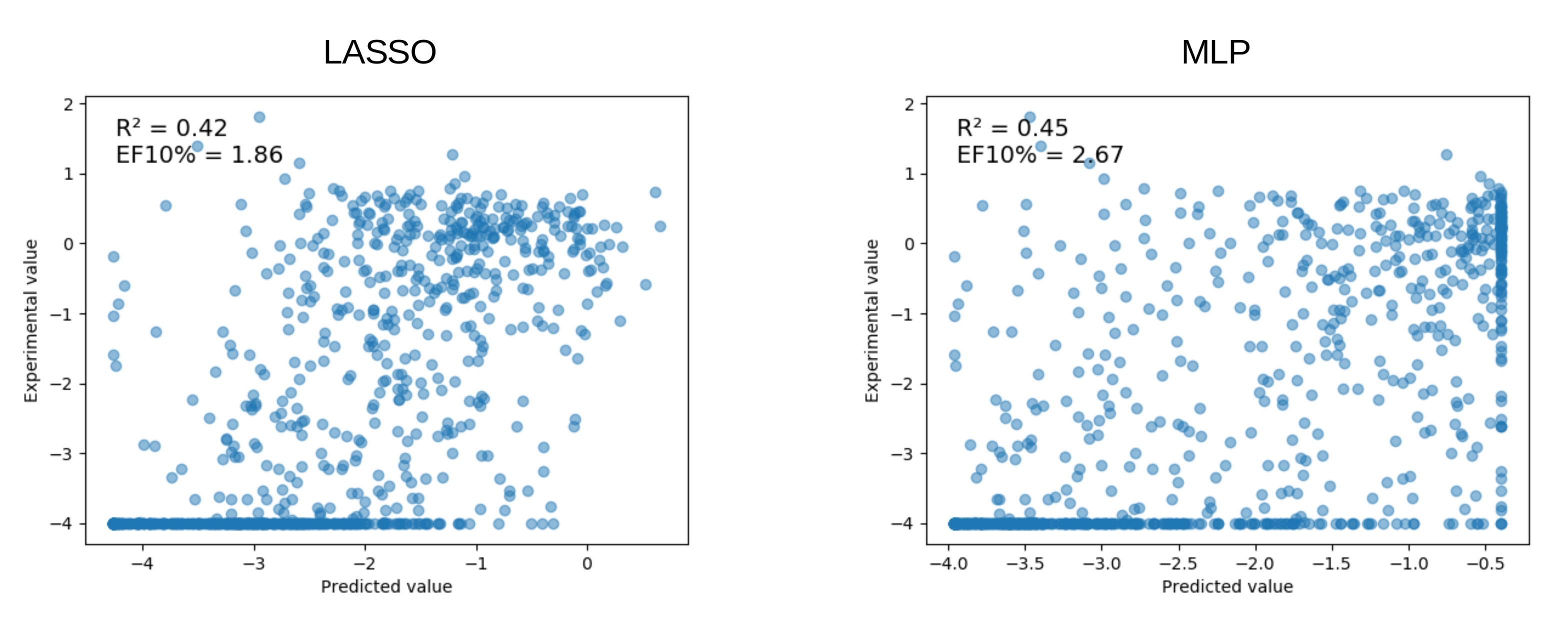}
    \caption{\textbf{Testing performance using only the solvent accessible area change and folding $\Delta\Delta$G.}}\label{fig:asa_ddg} 
\end{figure}

    \FloatBarrier

\begin{table}
    \caption{Testing and training R$^2$ are reported for LASSO and MLP models trained on the combination of Dynamical Signatures and all static descriptors (All), the Dynamical Signatures only (DynaSig), the static descriptors only (Static) and for the combination of folding $\Delta\Delta$G and accessible solvent area (ASA $\Delta\Delta$G).}\label{tab:r2}
    \begin{tabular}{ p{3cm}p{2cm}p{2cm}p{2cm}p{1.5cm}}
\hline
Training variables & ML model & Testing R$^2$ & Training R$^2$ & EF10\% \\
\hline
All & LASSO & 0.61 & 0.64 & 2.79\\
All & MLP & 0.69 & 0.96 & 3.49\\
DynaSig & LASSO & 0.24 & 0.29 & 2.79\\
DynaSig & MLP & 0.46 & 0.82 & 3.26\\
Static & LASSO & 0.52 & 0.54 & 2.79\\
Static & MLP & 0.58 & 0.78 & 3.26\\
ASA ddG & LASSO & 0.42 & 0.45 & 1.86\\
ASA ddG & MLP & 0.45 & 0.50 & 2.67\\
\hline
\end{tabular}

\end{table}

\FloatBarrier

    \begin{longtable}[c]{p{1cm}p{1.2cm}p{1cm}p{1cm}p{1.5cm}p{1.4cm}p{1cm}p{1cm}p{1.5cm}p{1.5cm}}
\caption{Top 10\% testing set predicted single mutants across all tested models. LASSO and MLP top 10\% predictions are reported for the combination of Dynamical Signatures and all static descriptors (all), for the Dynamical Signatures only (DynaSig), for the static descriptors only (static) and for the combination of folding $\Delta\Delta$G and accessible solvent area ($\Delta\Delta$G ASA). The true top 10\% experimental measurements for the testing set are identified in the second column.\label{tab:top10}}\\
\hline
Variant ID & Exp top 10\% & LASSO all & MLP all & LASSO DynaSig & MLP \newline DynaSig & LASSO static & MLP static & LASSO \newline ASA $\Delta\Delta$G & MLP \newline ASA $\Delta\Delta$G \\
\hline
\endfirsthead
\hline
\multicolumn{10}{ c }{Supplementary Table 4 (continued)}\\
\hline
Variant ID & Exp top 10\% & LASSO all & MLP all & LASSO DynaSig & MLP \newline DynaSig & LASSO static & MLP static & LASSO \newline ASA $\Delta\Delta$G & MLP \newline ASA $\Delta\Delta$G \\
\hline
\endhead
E$32$A & FALSE & IN & IN & OUT & OUT & IN & OUT & IN & IN\\
E$32$I & FALSE & IN & OUT & OUT & OUT & IN & OUT & IN & IN\\
E$32$M & FALSE & IN & OUT & OUT & OUT & IN & OUT & IN & IN\\
P$34$F & FALSE & OUT & OUT & IN & OUT & OUT & OUT & OUT & OUT\\
T$35$E & TRUE & OUT & OUT & IN & IN & OUT & OUT & OUT & OUT\\
T$35$I & TRUE & OUT & IN & OUT & IN & OUT & OUT & OUT & OUT\\
V$36$D & TRUE & OUT & OUT & IN & OUT & OUT & OUT & OUT & OUT\\
V$36$G & TRUE & IN & IN & IN & IN & OUT & OUT & OUT & OUT\\
V$36$H & TRUE & OUT & OUT & OUT & IN & OUT & OUT & OUT & OUT\\
S$37$D & FALSE & IN & IN & OUT & OUT & IN & OUT & IN & IN\\
S$37$E & FALSE & IN & OUT & OUT & OUT & IN & OUT & IN & OUT\\
S$37$M & FALSE & IN & OUT & OUT & OUT & IN & OUT & IN & IN\\
S$37$Q & FALSE & IN & OUT & OUT & OUT & IN & OUT & IN & IN\\
E$38$C & FALSE & OUT & OUT & OUT & IN & OUT & OUT & OUT & IN\\
E$38$H & FALSE & OUT & IN & OUT & IN & OUT & IN & OUT & OUT\\
E$38$R & FALSE & OUT & IN & OUT & IN & OUT & IN & OUT & OUT\\
V$41$G & TRUE & IN & OUT & IN & IN & OUT & OUT & OUT & OUT\\
V$41$I & FALSE & OUT & OUT & OUT & OUT & OUT & IN & OUT & IN\\
V$41$K & FALSE & OUT & OUT & OUT & OUT & OUT & IN & OUT & OUT\\
V$41$L & FALSE & OUT & OUT & OUT & OUT & OUT & IN & OUT & OUT\\
V$41$M & FALSE & OUT & IN & OUT & OUT & OUT & IN & OUT & OUT\\
V$41$P & TRUE & OUT & OUT & OUT & OUT & OUT & IN & OUT & OUT\\
V$41$R & FALSE & IN & OUT & OUT & OUT & OUT & IN & OUT & OUT\\
E$43$A & FALSE & OUT & IN & OUT & OUT & OUT & IN & OUT & OUT\\
E$43$H & FALSE & OUT & OUT & OUT & OUT & OUT & IN & OUT & OUT\\
E$43$I & FALSE & OUT & OUT & OUT & OUT & OUT & OUT & OUT & IN\\
E$43$K & FALSE & OUT & IN & IN & OUT & OUT & OUT & OUT & IN\\
R$45$E & FALSE & OUT & OUT & IN & OUT & OUT & OUT & OUT & OUT\\
R$45$M & TRUE & OUT & OUT & IN & OUT & OUT & OUT & OUT & OUT\\
Y$47$A & TRUE & OUT & OUT & IN & OUT & OUT & OUT & OUT & OUT\\
Y$47$G & FALSE & OUT & OUT & IN & OUT & OUT & OUT & OUT & OUT\\
Q$48$G & FALSE & OUT & OUT & OUT & IN & OUT & OUT & OUT & OUT\\
Q$48$H & FALSE & OUT & OUT & OUT & OUT & IN & OUT & OUT & IN\\
Q$48$L & FALSE & IN & OUT & OUT & IN & IN & OUT & OUT & IN\\
Q$48$P & TRUE & IN & OUT & OUT & OUT & IN & OUT & IN & IN\\
Q$48$R & TRUE & IN & OUT & IN & OUT & IN & IN & OUT & IN\\
A$50$K & FALSE & OUT & OUT & OUT & OUT & OUT & IN & OUT & OUT\\
A$50$Q & FALSE & OUT & OUT & OUT & OUT & OUT & IN & OUT & OUT\\
A$50$S & FALSE & OUT & IN & OUT & OUT & OUT & IN & OUT & OUT\\
D$51$E & FALSE & OUT & OUT & OUT & OUT & OUT & OUT & IN & IN\\
D$51$I & FALSE & OUT & OUT & OUT & OUT & OUT & OUT & IN & OUT\\
D$51$L & FALSE & OUT & OUT & OUT & OUT & OUT & OUT & IN & IN\\
D$51$R & FALSE & OUT & OUT & OUT & OUT & OUT & OUT & IN & OUT\\
D$51$V & FALSE & OUT & OUT & OUT & OUT & IN & OUT & IN & IN\\
S$55$A & TRUE & OUT & IN & IN & IN & OUT & OUT & OUT & OUT\\
S$55$F & FALSE & OUT & OUT & IN & OUT & OUT & OUT & OUT & OUT\\
S$55$M & TRUE & OUT & IN & IN & IN & OUT & OUT & OUT & OUT\\
S$55$P & FALSE & OUT & OUT & IN & OUT & OUT & OUT & OUT & OUT\\
S$61$F & FALSE & OUT & OUT & OUT & OUT & OUT & OUT & OUT & IN\\
S$61$N & FALSE & IN & OUT & OUT & OUT & IN & IN & IN & OUT\\
D$63$G & TRUE & OUT & OUT & OUT & OUT & IN & OUT & IN & OUT\\
D$63$I & FALSE & OUT & OUT & OUT & OUT & OUT & OUT & IN & OUT\\
D$63$L & FALSE & OUT & OUT & OUT & OUT & IN & OUT & IN & IN\\
D$63$N & FALSE & IN & OUT & OUT & OUT & IN & IN & IN & OUT\\
D$63$W & FALSE & OUT & OUT & OUT & OUT & OUT & OUT & IN & IN\\
G$64$E & FALSE & OUT & OUT & OUT & OUT & OUT & OUT & IN & OUT\\
A$65$G & FALSE & OUT & IN & OUT & OUT & OUT & OUT & OUT & OUT\\
A$65$H & TRUE & OUT & IN & OUT & OUT & OUT & OUT & OUT & OUT\\
A$65$Y & TRUE & OUT & IN & OUT & OUT & OUT & IN & OUT & OUT\\
V$66$F & TRUE & OUT & IN & OUT & OUT & OUT & OUT & OUT & OUT\\
V$66$R & TRUE & OUT & OUT & OUT & IN & OUT & IN & OUT & OUT\\
V$66$Y & TRUE & OUT & OUT & OUT & OUT & OUT & IN & OUT & OUT\\
Y$67$A & TRUE & OUT & OUT & IN & OUT & OUT & OUT & OUT & OUT\\
Y$67$H & FALSE & OUT & IN & OUT & OUT & OUT & IN & OUT & OUT\\
P$68$C & FALSE & OUT & OUT & IN & OUT & OUT & OUT & OUT & OUT\\
P$68$Y & TRUE & OUT & OUT & IN & IN & OUT & OUT & OUT & OUT\\
D$78$C & FALSE & OUT & OUT & OUT & OUT & OUT & OUT & IN & OUT\\
D$78$M & FALSE & OUT & OUT & OUT & OUT & OUT & OUT & IN & OUT\\
D$78$P & FALSE & OUT & OUT & OUT & OUT & OUT & OUT & IN & OUT\\
D$78$Q & FALSE & OUT & OUT & OUT & OUT & OUT & OUT & IN & OUT\\
D$78$V & FALSE & OUT & OUT & OUT & OUT & OUT & OUT & IN & OUT\\
D$78$W & FALSE & OUT & OUT & OUT & IN & OUT & OUT & IN & OUT\\
E$79$C & FALSE & OUT & OUT & IN & OUT & OUT & OUT & OUT & OUT\\
E$79$G & FALSE & OUT & OUT & IN & OUT & OUT & OUT & OUT & IN\\
I$83$F & FALSE & OUT & OUT & IN & OUT & OUT & OUT & OUT & OUT\\
A$89$H & FALSE & OUT & IN & OUT & OUT & OUT & OUT & OUT & OUT\\
K$90$A & TRUE & IN & IN & OUT & IN & IN & IN & OUT & OUT\\
A$93$L & TRUE & OUT & IN & OUT & IN & IN & OUT & OUT & OUT\\
A$93$T & TRUE & OUT & IN & OUT & OUT & OUT & IN & OUT & OUT\\
A$94$E & FALSE & OUT & IN & OUT & OUT & OUT & OUT & OUT & OUT\\
A$97$I & FALSE & IN & OUT & OUT & OUT & IN & OUT & OUT & OUT\\
A$97$Q & FALSE & IN & IN & OUT & OUT & IN & IN & OUT & IN\\
E$98$Y & FALSE & OUT & OUT & IN & OUT & OUT & OUT & OUT & OUT\\
I$99$W & FALSE & OUT & OUT & IN & OUT & OUT & OUT & OUT & OUT\\
E$100$M & FALSE & IN & IN & OUT & OUT & IN & IN & IN & OUT\\
E$100$R & TRUE & OUT & IN & OUT & OUT & OUT & IN & OUT & OUT\\
E$100$W & FALSE & OUT & IN & OUT & OUT & OUT & IN & IN & IN\\
K$101$F & FALSE & IN & OUT & OUT & IN & IN & OUT & IN & OUT\\
K$101$Y & FALSE & IN & OUT & OUT & IN & IN & OUT & IN & OUT\\
Q$102$H & TRUE & IN & OUT & IN & OUT & OUT & OUT & OUT & OUT\\
Q$102$K & TRUE & OUT & IN & OUT & IN & OUT & IN & OUT & OUT\\
Q$102$L & TRUE & IN & IN & OUT & IN & IN & OUT & OUT & IN\\
Q$102$T & TRUE & IN & IN & OUT & OUT & OUT & IN & OUT & OUT\\
L$105$A & FALSE & OUT & OUT & IN & OUT & OUT & OUT & OUT & OUT\\
L$105$S & FALSE & OUT & OUT & IN & OUT & OUT & OUT & OUT & OUT\\
R$109$A & FALSE & OUT & IN & OUT & OUT & OUT & OUT & OUT & OUT\\
T$113$M & FALSE & OUT & OUT & IN & OUT & OUT & OUT & OUT & OUT\\
H$114$E & FALSE & OUT & OUT & OUT & OUT & OUT & OUT & IN & OUT\\
F$115$Y & FALSE & OUT & OUT & IN & OUT & OUT & OUT & OUT & OUT\\
D$117$L & FALSE & OUT & OUT & OUT & OUT & OUT & OUT & OUT & IN\\
D$117$N & FALSE & OUT & OUT & OUT & OUT & OUT & OUT & OUT & IN\\
D$117$V & FALSE & OUT & OUT & OUT & OUT & OUT & OUT & OUT & IN\\
G$121$D & FALSE & OUT & IN & OUT & OUT & OUT & OUT & OUT & OUT\\
D$124$F & FALSE & OUT & OUT & OUT & OUT & OUT & OUT & OUT & IN\\
D$124$N & FALSE & OUT & IN & OUT & OUT & OUT & OUT & IN & IN\\
V$125$M & FALSE & OUT & IN & OUT & OUT & OUT & IN & OUT & OUT\\
R$127$E & FALSE & OUT & OUT & IN & IN & OUT & OUT & OUT & OUT\\
R$127$T & FALSE & OUT & OUT & OUT & IN & OUT & OUT & OUT & OUT\\
A$128$C & FALSE & OUT & OUT & OUT & OUT & OUT & OUT & IN & OUT\\
A$128$R & FALSE & IN & OUT & OUT & OUT & IN & OUT & IN & IN\\
A$129$C & FALSE & OUT & OUT & OUT & IN & OUT & OUT & OUT & OUT\\
A$129$E & FALSE & OUT & OUT & OUT & IN & OUT & IN & OUT & OUT\\
A$129$G & FALSE & IN & IN & OUT & OUT & OUT & IN & OUT & OUT\\
A$129$S & FALSE & IN & IN & OUT & OUT & OUT & IN & OUT & OUT\\
A$129$V & FALSE & OUT & OUT & OUT & IN & OUT & OUT & OUT & OUT\\
A$132$K & TRUE & OUT & IN & OUT & OUT & OUT & IN & OUT & OUT\\
A$132$M & TRUE & OUT & IN & OUT & OUT & OUT & OUT & OUT & OUT\\
A$132$N & FALSE & IN & IN & OUT & OUT & IN & IN & OUT & OUT\\
P$137$K & FALSE & OUT & OUT & OUT & IN & OUT & OUT & OUT & OUT\\
S$138$C & FALSE & OUT & OUT & OUT & OUT & OUT & OUT & OUT & IN\\
S$138$K & FALSE & OUT & OUT & IN & OUT & OUT & OUT & OUT & OUT\\
S$138$L & FALSE & OUT & OUT & OUT & OUT & OUT & OUT & OUT & IN\\
S$138$T & FALSE & IN & OUT & OUT & OUT & OUT & OUT & OUT & IN\\
S$138$Y & FALSE & OUT & OUT & IN & OUT & OUT & OUT & OUT & OUT\\
R$141$H & TRUE & OUT & OUT & OUT & OUT & IN & OUT & IN & IN\\
R$141$K & TRUE & IN & OUT & OUT & OUT & IN & IN & IN & OUT\\
R$141$M & FALSE & IN & IN & OUT & OUT & IN & IN & IN & IN\\
E$144$K & TRUE & IN & IN & OUT & OUT & IN & IN & IN & IN\\
E$144$V & FALSE & OUT & OUT & OUT & OUT & IN & OUT & OUT & OUT\\
V$145$D & TRUE & OUT & IN & OUT & IN & OUT & IN & IN & IN\\
V$145$K & TRUE & IN & OUT & OUT & IN & IN & IN & IN & IN\\
E$146$A & TRUE & OUT & IN & OUT & IN & OUT & IN & IN & IN\\
E$146$I & FALSE & OUT & OUT & IN & OUT & OUT & OUT & OUT & OUT\\
E$146$R & FALSE & IN & IN & OUT & IN & OUT & IN & OUT & OUT\\
G$147$N & FALSE & OUT & OUT & OUT & OUT & OUT & OUT & IN & OUT\\
N$148$E & FALSE & OUT & OUT & OUT & IN & OUT & OUT & OUT & OUT\\
N$148$M & TRUE & OUT & IN & OUT & OUT & OUT & OUT & OUT & OUT\\
E$149$R & FALSE & OUT & OUT & IN & OUT & OUT & OUT & OUT & OUT\\
I$150$V & FALSE & OUT & OUT & OUT & OUT & OUT & IN & OUT & OUT\\
H$153$P & FALSE & OUT & OUT & IN & OUT & OUT & OUT & OUT & OUT\\
H$153$S & FALSE & OUT & OUT & IN & OUT & OUT & OUT & OUT & OUT\\
S$154$A & TRUE & OUT & IN & OUT & OUT & OUT & IN & OUT & OUT\\
S$154$D & FALSE & OUT & OUT & OUT & OUT & OUT & OUT & OUT & IN\\
E$156$I & FALSE & OUT & OUT & OUT & IN & IN & OUT & IN & IN\\
E$156$P & FALSE & IN & OUT & OUT & IN & IN & OUT & IN & IN\\
E$156$S & FALSE & IN & OUT & OUT & IN & IN & IN & IN & IN\\
S$159$D & TRUE & IN & OUT & IN & OUT & IN & OUT & IN & IN\\
S$159$E & TRUE & IN & OUT & IN & OUT & IN & OUT & IN & IN\\
S$159$N & FALSE & IN & IN & OUT & OUT & IN & OUT & IN & IN\\
S$160$D & FALSE & OUT & IN & OUT & OUT & OUT & IN & IN & IN\\
S$160$K & TRUE & IN & OUT & OUT & IN & IN & OUT & IN & IN\\
S$161$W & FALSE & OUT & IN & OUT & IN & OUT & OUT & OUT & OUT\\
A$164$D & FALSE & OUT & OUT & OUT & IN & OUT & OUT & OUT & OUT\\
A$164$I & FALSE & OUT & OUT & OUT & IN & OUT & OUT & OUT & OUT\\
A$164$L & TRUE & OUT & OUT & OUT & IN & OUT & OUT & OUT & OUT\\
A$164$M & TRUE & OUT & IN & IN & IN & OUT & OUT & OUT & OUT\\
A$164$Q & FALSE & IN & IN & OUT & IN & OUT & IN & OUT & OUT\\
V$165$I & FALSE & OUT & OUT & OUT & OUT & OUT & IN & OUT & OUT\\
V$165$M & FALSE & OUT & OUT & OUT & OUT & OUT & IN & OUT & OUT\\
R$166$Q & TRUE & IN & IN & IN & IN & OUT & IN & OUT & IN\\
R$166$V & FALSE & IN & IN & IN & IN & IN & OUT & IN & IN\\
V$170$F & FALSE & OUT & OUT & IN & OUT & OUT & OUT & OUT & OUT\\
V$170$M & FALSE & OUT & OUT & IN & OUT & OUT & OUT & OUT & OUT\\
A$177$L & FALSE & OUT & OUT & OUT & IN & OUT & OUT & OUT & OUT\\
A$177$Q & FALSE & OUT & OUT & IN & OUT & OUT & OUT & OUT & OUT\\
A$177$Y & FALSE & OUT & OUT & OUT & IN & OUT & OUT & OUT & OUT\\
A$178$G & FALSE & OUT & OUT & IN & IN & OUT & OUT & OUT & OUT\\
T$181$D & FALSE & IN & OUT & IN & OUT & OUT & IN & OUT & OUT\\
T$181$E & FALSE & OUT & IN & OUT & IN & OUT & IN & OUT & IN\\
T$181$H & FALSE & OUT & IN & OUT & IN & OUT & IN & OUT & OUT\\
T$181$L & TRUE & OUT & OUT & OUT & OUT & IN & OUT & OUT & IN\\
T$181$R & FALSE & IN & IN & OUT & IN & IN & OUT & OUT & IN\\
T$181$V & FALSE & IN & OUT & OUT & OUT & IN & IN & OUT & OUT\\
T$181$Y & FALSE & IN & OUT & IN & OUT & OUT & OUT & OUT & OUT\\
L$184$M & FALSE & OUT & OUT & IN & OUT & OUT & OUT & OUT & OUT\\
I$185$V & TRUE & OUT & OUT & OUT & OUT & OUT & IN & OUT & OUT\\
S$190$D & FALSE & IN & OUT & OUT & OUT & IN & OUT & IN & IN\\
S$190$E & FALSE & IN & OUT & OUT & OUT & IN & OUT & IN & IN\\
S$190$H & TRUE & IN & OUT & OUT & IN & IN & OUT & IN & OUT\\
S$190$K & FALSE & IN & OUT & OUT & OUT & IN & OUT & IN & OUT\\
S$190$N & FALSE & IN & OUT & OUT & OUT & IN & OUT & IN & IN\\
S$190$Y & FALSE & IN & OUT & OUT & OUT & IN & OUT & IN & IN\\
S$192$A & TRUE & OUT & IN & OUT & IN & OUT & IN & OUT & OUT\\
S$192$H & FALSE & OUT & IN & OUT & OUT & OUT & OUT & OUT & OUT\\
S$192$N & FALSE & OUT & OUT & OUT & OUT & OUT & IN & OUT & OUT\\
S$192$T & FALSE & OUT & OUT & OUT & OUT & OUT & IN & OUT & OUT\\
V$193$K & FALSE & OUT & OUT & OUT & IN & OUT & OUT & OUT & OUT\\
V$193$R & FALSE & OUT & OUT & OUT & IN & OUT & OUT & OUT & OUT\\
A$199$C & FALSE & OUT & OUT & IN & OUT & OUT & OUT & OUT & OUT\\
A$199$F & FALSE & OUT & OUT & IN & OUT & OUT & OUT & OUT & OUT\\
I$200$V & TRUE & OUT & IN & OUT & OUT & OUT & IN & OUT & OUT\\
L$203$F & FALSE & OUT & OUT & IN & OUT & OUT & OUT & OUT & OUT\\
L$203$G & FALSE & OUT & OUT & IN & IN & OUT & OUT & OUT & OUT\\
S$204$D & FALSE & IN & IN & OUT & OUT & IN & OUT & IN & IN\\
S$204$E & TRUE & OUT & OUT & OUT & OUT & IN & OUT & IN & IN\\
S$204$F & FALSE & OUT & OUT & OUT & OUT & IN & IN & IN & OUT\\
S$204$H & FALSE & OUT & OUT & OUT & OUT & IN & OUT & IN & IN\\
S$204$T & FALSE & IN & OUT & OUT & OUT & IN & OUT & IN & IN\\
R$205$D & FALSE & OUT & OUT & IN & OUT & OUT & OUT & OUT & OUT\\
R$205$F & FALSE & OUT & OUT & IN & OUT & OUT & OUT & OUT & OUT\\
T$206$A & FALSE & IN & IN & IN & IN & IN & IN & IN & OUT\\
T$206$C & FALSE & OUT & OUT & IN & IN & OUT & OUT & IN & IN\\
T$206$K & TRUE & IN & IN & IN & IN & IN & IN & IN & IN\\
T$206$N & TRUE & IN & IN & OUT & IN & IN & OUT & IN & IN\\
T$206$R & FALSE & IN & IN & IN & IN & IN & IN & IN & OUT\\
S$207$T & TRUE & OUT & OUT & OUT & OUT & OUT & IN & OUT & OUT\\
N$210$E & FALSE & IN & OUT & OUT & OUT & IN & IN & IN & IN\\
N$210$L & FALSE & OUT & OUT & OUT & OUT & OUT & IN & OUT & OUT\\
A$212$T & FALSE & IN & OUT & OUT & OUT & IN & IN & IN & IN\\
D$213$N & FALSE & OUT & OUT & OUT & IN & OUT & OUT & OUT & OUT\\
A$217$C & FALSE & OUT & OUT & OUT & OUT & OUT & OUT & IN & IN\\
A$217$N & FALSE & IN & IN & OUT & OUT & IN & OUT & IN & IN\\
A$217$Y & TRUE & IN & OUT & OUT & OUT & IN & OUT & OUT & OUT\\
E$218$Q & FALSE & OUT & OUT & OUT & OUT & OUT & OUT & OUT & IN\\
T$221$D & FALSE & OUT & IN & OUT & OUT & OUT & IN & OUT & IN\\
T$221$L & FALSE & IN & IN & OUT & OUT & IN & OUT & OUT & IN\\
T$221$Q & TRUE & OUT & IN & OUT & OUT & IN & IN & OUT & IN\\
T$221$S & TRUE & OUT & IN & OUT & OUT & IN & OUT & OUT & OUT\\
S$222$G & FALSE & OUT & OUT & IN & IN & OUT & OUT & OUT & OUT\\
S$222$N & FALSE & OUT & OUT & OUT & IN & OUT & OUT & OUT & OUT\\
S$222$P & FALSE & OUT & OUT & IN & IN & OUT & OUT & OUT & OUT\\
I$223$F & FALSE & OUT & OUT & IN & OUT & OUT & OUT & OUT & OUT\\
I$223$V & FALSE & OUT & OUT & OUT & OUT & OUT & IN & OUT & OUT\\
E$224$G & FALSE & OUT & IN & OUT & OUT & OUT & OUT & OUT & OUT\\
R$225$A & FALSE & OUT & OUT & IN & OUT & OUT & OUT & OUT & OUT\\
R$225$G & FALSE & OUT & OUT & IN & OUT & OUT & OUT & OUT & OUT\\
Q$227$I & FALSE & IN & OUT & OUT & OUT & OUT & OUT & OUT & OUT\\
Q$227$L & FALSE & OUT & OUT & IN & OUT & OUT & OUT & OUT & OUT\\
Q$227$M & TRUE & OUT & OUT & IN & OUT & OUT & IN & OUT & OUT\\
Q$228$D & FALSE & IN & IN & OUT & IN & IN & IN & IN & IN\\
Q$228$E & FALSE & IN & IN & OUT & IN & IN & OUT & IN & IN\\
Q$228$G & FALSE & OUT & OUT & OUT & IN & OUT & OUT & IN & OUT\\
Q$228$K & TRUE & IN & IN & OUT & OUT & IN & OUT & IN & IN\\
Q$228$V & FALSE & IN & OUT & OUT & IN & IN & IN & OUT & OUT\\
H$229$A & TRUE & OUT & IN & OUT & IN & OUT & IN & OUT & OUT\\
H$229$G & FALSE & OUT & OUT & OUT & IN & OUT & OUT & OUT & IN\\
H$229$I & FALSE & OUT & IN & OUT & OUT & OUT & OUT & OUT & IN\\
H$229$L & FALSE & OUT & OUT & IN & OUT & OUT & OUT & OUT & OUT\\
H$229$P & FALSE & OUT & OUT & OUT & IN & OUT & OUT & OUT & OUT\\
H$229$Q & FALSE & OUT & OUT & OUT & OUT & OUT & OUT & IN & OUT\\
H$229$V & FALSE & OUT & OUT & OUT & OUT & OUT & IN & OUT & OUT\\
P$231$H & FALSE & OUT & IN & OUT & OUT & OUT & OUT & OUT & OUT\\
P$231$T & FALSE & OUT & IN & OUT & OUT & OUT & IN & OUT & OUT\\
Q$234$C & FALSE & OUT & OUT & OUT & OUT & OUT & OUT & IN & OUT\\
Q$234$E & FALSE & IN & OUT & OUT & OUT & IN & OUT & IN & OUT\\
Q$234$N & FALSE & IN & OUT & OUT & OUT & IN & OUT & IN & OUT\\
Q$234$R & FALSE & IN & OUT & OUT & OUT & IN & IN & IN & IN\\
Q$234$S & FALSE & IN & OUT & OUT & OUT & IN & IN & IN & OUT\\
F$235$W & FALSE & OUT & OUT & IN & OUT & OUT & OUT & OUT & OUT\\
L$242$D & TRUE & IN & OUT & IN & IN & OUT & OUT & OUT & OUT\\
L$242$F & FALSE & OUT & OUT & IN & OUT & OUT & OUT & OUT & OUT\\
L$242$K & TRUE & OUT & OUT & IN & IN & OUT & OUT & OUT & OUT\\
L$242$V & TRUE & IN & IN & OUT & OUT & OUT & IN & OUT & IN\\
L$246$D & FALSE & IN & IN & IN & IN & OUT & OUT & OUT & OUT\\
L$246$G & FALSE & OUT & OUT & IN & IN & OUT & OUT & OUT & OUT\\
L$246$Q & FALSE & IN & IN & IN & IN & OUT & OUT & OUT & OUT\\
L$246$S & TRUE & OUT & OUT & IN & OUT & OUT & OUT & OUT & OUT\\
L$246$T & TRUE & IN & OUT & IN & OUT & OUT & OUT & OUT & OUT\\
D$247$L & FALSE & OUT & OUT & OUT & OUT & OUT & OUT & IN & IN\\
D$247$M & FALSE & OUT & OUT & OUT & OUT & OUT & OUT & IN & OUT\\
D$247$R & FALSE & OUT & OUT & OUT & OUT & OUT & OUT & IN & IN\\
D$247$Y & FALSE & OUT & OUT & OUT & OUT & OUT & OUT & IN & IN\\
L$248$Y & FALSE & OUT & OUT & IN & OUT & OUT & OUT & OUT & OUT\\
K$250$E & FALSE & OUT & IN & OUT & IN & IN & IN & OUT & OUT\\
K$250$F & FALSE & OUT & OUT & IN & OUT & IN & OUT & OUT & OUT\\
K$250$N & FALSE & IN & IN & OUT & IN & IN & IN & OUT & IN\\
T$253$I & FALSE & OUT & IN & OUT & OUT & OUT & OUT & OUT & OUT\\
N$254$D & FALSE & OUT & OUT & OUT & IN & OUT & OUT & OUT & OUT\\
N$254$G & TRUE & OUT & OUT & IN & OUT & OUT & OUT & OUT & OUT\\
N$254$I & FALSE & OUT & OUT & IN & IN & OUT & OUT & OUT & IN\\
N$254$P & FALSE & OUT & OUT & IN & OUT & OUT & OUT & OUT & OUT\\
K$257$G & FALSE & OUT & OUT & IN & OUT & OUT & OUT & OUT & OUT\\
K$257$N & FALSE & OUT & OUT & IN & IN & IN & IN & OUT & OUT\\
K$257$Q & TRUE & IN & OUT & IN & IN & IN & IN & OUT & OUT\\
K$257$T & FALSE & OUT & OUT & OUT & OUT & IN & OUT & OUT & OUT\\
K$257$V & FALSE & IN & OUT & IN & OUT & IN & IN & OUT & OUT\\
A$258$F & FALSE & OUT & OUT & OUT & OUT & IN & OUT & OUT & OUT\\
A$258$V & FALSE & IN & IN & OUT & OUT & IN & IN & OUT & OUT\\
H$259$I & FALSE & OUT & IN & OUT & OUT & OUT & OUT & OUT & OUT\\
H$259$S & FALSE & IN & OUT & IN & IN & OUT & OUT & OUT & OUT\\
T$260$D & FALSE & OUT & OUT & OUT & IN & OUT & IN & OUT & OUT\\
T$260$K & FALSE & OUT & IN & IN & IN & IN & OUT & OUT & IN\\
T$260$M & FALSE & IN & OUT & IN & IN & IN & IN & OUT & OUT\\
T$260$R & FALSE & IN & IN & IN & OUT & IN & OUT & OUT & IN\\
N$261$H & FALSE & IN & IN & OUT & OUT & IN & OUT & IN & IN\\
R$262$K & TRUE & IN & OUT & OUT & IN & IN & IN & IN & OUT\\
\hline
\end{longtable}

    \bibliographystyle{unsrt}
    \bibliography{used_refs}